\documentclass[10pt,twocolumn]{IEEEtran}

\usepackage{amsmath}
\usepackage{amsfonts}
\usepackage{epsfig}
\usepackage{amssymb}
\usepackage[nospace]{cite}
\usepackage{color,soul}
\usepackage{subfigure}
\usepackage{multirow}
\usepackage{rotating}
\usepackage{graphicx}
\usepackage{tabularx}
\usepackage{array}
\newcolumntype{P}[1]{>{\centering\arraybackslash}p{#1}}
\newcolumntype{M}[1]{>{\centering\arraybackslash}m{#1}}
\usepackage{ragged2e}

\usepackage{makecell}
\usepackage{array}
\usepackage{blindtext}
\usepackage{ragged2e}
\usepackage{xcolor,colortbl}
\definecolor{Gray}{gray}{0.85}

\usepackage{amsthm,amssymb,amsmath,bm}
\usepackage{fancyhdr}
\usepackage[subfigure]{tocloft}
\usepackage[font={small}]{caption}
\usepackage{graphicx}
\usepackage[caption=false]{subfig}
\usepackage[utf8]{inputenc}
\usepackage[english]{babel}

\usepackage{lipsum} 
\usepackage{algorithm}
\usepackage{algpseudocode}
\usepackage{stackengine}
\def\delequal{\mathrel{\ensurestackMath{\stackon[1pt]{=}{\scriptstyle\Delta}}}}
\makeatletter
\newcommand{\algmargin}{\the\ALG@thistlm}
\makeatother
\newlength{\whilewidth}
\algdef{SE}[parWHILE]{parWhile}{EndparWhile}[1]
  {\parbox[t]{\dimexpr\linewidth-\algmargin}{%
     \hangindent\whilewidth\strut\algorithmicwhile\ #1\ \algorithmicdo\strut}}{\algorithmicend\ \algorithmicwhile}%
\algnewcommand{\parState}[1]{\State%
  \parbox[t]{\dimexpr\linewidth-\algmargin}{\strut #1\strut}}
\usepackage{tabularx}
\definecolor{Gray}{gray}{0.85}
\newtheorem{remark}{Remark}

\newcolumntype{C}[1]{>{\centering\arraybackslash}p{#1}}
\begin{document}
\vspace{-20mm}
\title{Physical Layer Group Key Generation With the Aid of Reconfigurable Intelligent Surfaces}
\author{Vahid Shahiri, Hamid Behroozi,~\IEEEmembership{Member,~IEEE} \\~~\\

\thanks{ABC}. 
}
\author{Vahid Shahiri, Guyue Li,~\IEEEmembership{Member,~IEEE} 
	\thanks{V. Shahiri and H. Behroozi are with the Electrical Engineering Department, Sharif		University of Technology, Tehran, Iran. E-mail: vahid.shahiri@ee.sharif.edu, behroozi@sharif.edu}
	\thanks{Guyue Li is with the School of Cyber Science and Engineering, Southeast University, Nanjing 210096, China, and also with the Purple Mountain Laboratories for Network and Communication Security, Nanjing 210096, China (e-mail: guyuelee@seu.edu.cn).},
	 Hamid Behroozi,~\IEEEmembership{Member,~IEEE}}
\maketitle
{\vspace{-10mm}}
\begin{abstract}
Reconfigurable intelligent surfaces (RIS) have the ability to alter the wireless environment by making changes in the impinging signal. While RIS has been extensively studied for enhancing wireless communications, its potential for facilitating group key generation (GKG) remains unexplored. Leveraging this capability, in this study, we exploit the RIS to make the aggregate reflecting channels of different user terminals (UTs) as similar as possible to be able to extract common group secret keys from their channels. Specifically, the RIS will adjust its parameters to pave the way for GKG based on the physical channels of the UTs. Our method exploits the already gathered channel state information (CSI) in the RIS to beneficially design the phase shifts and does not impose additional probing burden on the network. Additionally, this scheme is broadcast-based and does not entail the overheads of the pairwise-based key generation. We consider both passive RIS (PRIS) and active RIS (ARIS) to generate the group keys. The PRIS is widely adopted in physical layer key generation (PLKG) studies due to its use of passive elements, whereas the ARIS demonstrates superior capability in aligning the aggregate reflected channels among nodes in the GKG scenario, as demonstrated in this study. We will exploit various optimization methods like successive convex approximation (SCA) and semidefinite relaxation with Gaussian randomization (SDR-GR) to address the raised optimization problems. Unlike most of the studies in the literature, our scheme can achieve a high GKG rate in static environments as well. Finally, we will examine the performance of the proposed method by normalized mean squared error (NMSE), key error rate (KER), key generation rate (KGR) and key randomness metrics. Our numerical results verify that for the equal available power budget, the ARIS significantly outperforms PRIS in NMSE and KER, achieving more than four times higher KGR.
\end{abstract}
\IEEEpeerreviewmaketitle
\section{Introduction}
\IEEEPARstart{S}{ecurity} remains a critical challenge in the deployment of the Internet of Things (IoT) networks. The inherent broadcast nature of the wireless medium makes it susceptible to wiretapping. Meanwhile, the resource-constrained IoT devices cannot afford the complexity of the conventional cryptographic schemes, i.e., public key cryptography for the key distribution \cite{frontier_Access_2020}. Physical layer key generation (PLKG) offers a promising alternative by exploiting the reciprocity of the wireless channel between the nodes to derive the identical keys with minimal computational overhead. These keys can subsequently be deployed in the symmetric key encryption, providing a lightweight yet secure solution for the IoT networks \cite{TIFs_2024_multipath}.   

Reconfigurable intelligent surfaces (RIS) have emerged as a promising enabler for the PLKG due to their ability to dynamically modify the wireless environment. Recent studies have investigated the impact of diverse RIS structures on PLKG performance, including passive RIS (PRIS) \cite{JIoT-2023-Vahid,LWC-2023-LeiHu,TIFs-2022-Guyue,TWC-2023-LeiHu}, active RIS (ARIS) \cite{LWC_2023_ARIS}, simultaneously transmitting and reflecting RIS (STAR-RIS) \cite{STAR-RIS_TVT_2024_1,STAR-RIS_TVT_2024_2,STAR-RIS_TIFs_2024,STAR-RIS_JIoT_2024} and beyond-diagonal RIS (BD-RIS) \cite{SPAWC-2024}. Specifically, in \cite{JIoT-2023-Vahid} we proposed a novel random phase shift configuration for spatially correlated PRIS elements to enhance the key generation rate (KGR) in the quasi-static channels. Meanwhile, studies such as \cite{TIFs-2022-Guyue},\cite{TWC-2023-LeiHu} and \cite{LWC_2023_ARIS,STAR-RIS_TVT_2024_1,STAR-RIS_TVT_2024_2,STAR-RIS_TIFs_2024,STAR-RIS_JIoT_2024} focused on optimizing the beamforming at the RIS or the multiple-input multiple-output (MIMO) transceiver to enhance KGR. In \cite{LWC-2023-LeiHu} and \cite{SPAWC-2024} the authors have considered RIS as an adversary trying to disrupt the channel reciprocity and degrade PLKG performance. However, none of these studies have considered RIS as an enabler for the group key generation (GKG). We note that the PRIS remains the most widely considered RIS due to its low-cost, energy-efficient passive elements. In contrast, ARIS mitigates the double-fading effect inherent to PRIS \cite{ARIS_TWC_2021} and offers greater control over the wireless medium. Against this background and because these two are more representative, for simplicity, this study focuses on PRIS and ARIS-aided GKG, leaving STAR-RIS and BD-RIS to future studies.

Current research predominantly focuses on the pairwise PLKG between two legitimate parties. However, many practical applications require group key agreement, where multiple nodes must derive identical keys to share confidential information with each other. For instance, a group of police officers may need to share confidential information about a crime scene using the group key \cite{TIFs_2018_meshQuek}. Similarly, in IoT-enabled smart homes, an access point (AP) should be able to securely share the environment monitoring information with several automated home IoT devices \cite{JIoT_2022_Broadcast_Group}. Extending the pairwise key generation scenarios to group use cases is not straightforward, as the random channels vary across spatially distributed users. To address this challenge, numerous studies have striven to bring about solutions for the GKG problem. 

GKG mechanisms can broadly classified into two categories. The first category involves algorithms that generate group keys through pairwise key establishment \cite{TIFs_2016_group,LWC_2020_acoustic,CNS_2020_efficient,OJCOMS_2020_selforg}. As the group size increases, these methods require an impractical number of timeslots to generate the common keys. Accordingly, such approaches are primarily suitable for the networks with limited number of nodes \cite{JSYST_2022}. We note that the probing complexity for the methods entailing pairwise key generation between each of the nodes is $\mathcal{O}(K^2)$ where $K$ is the number of users within the group. This is because $\frac{K(K-1)}{2}\times 2$ rounds of probings is needed to generate pairwise keys. 

The second category of GKG relies primarily on broadcasting, making it well-suited for large scale networks \cite{JIoT_2022_Broadcast_Group}, \cite{LCOMM_2018_networkcoding,TVT_2019_OFDMA,ICCW_2019_lightweightGuyue}. Specifically, a precoding scheme at the AP is proposed in \cite{JIoT_2022_Broadcast_Group} enabling it to deactivate a specific antenna within the antennas of the nodes. The index of the deactivated antenna is the common key source. While having low complexity overhead, the KGR is constrained by the number of antennas. The authors in \cite{LCOMM_2018_networkcoding} employ a reference node and a central node, where channel information between them provides shared randomness, distributed via secure network coding (SNC). GKG in the orthogonal frequency-division multiplexing access (OFDMA) is considered in \cite{TVT_2019_OFDMA}. Exploiting the ability of OFDMA in simultaneously transmitting signals through different subcarriers, the authors have managed to efficiently generate private keys between the AP and each user. The common group key is eventually encrypted with these private keys and is sent to each user. Yet being efficient, the idea is only limited to OFDMA systems. A scalable lightweight GKG algorithm is proposed in \cite{ICCW_2019_lightweightGuyue}. The scalability is improved through omitting reconciliation phase in the pairwise key generation and the common randomness is shared through efficiently generated private keys with the AP. For the broadcast-based methods we can achieve $\mathcal{O}(K)$ complexity, significantly improving scalability.     

In this study, we investigate GKG in an RIS-assisted network comprising of multiple user terminals (UTs) and an AP. The RIS is primarily deployed to facilitate reliable communication in the harsh non-line-of-sight (NLOS) wireless environment, where direct links between the UTs and the AP are blocked \footnote{RIS is particularly beneficial when the direct link between the transmitter and receiver is blocked, as it enables the establishment of an alternative propagation path \cite{direct_block_5}. Accordingly, blockage of the direct link is a common assumption in RIS-assisted studies, both in general communication scenarios \cite{direct_block_1,direct_block_2,direct_block_3} and in PLKG applications \cite{TIFs-2022-Guyue}, \cite{STAR-RIS_TVT_2024_1}, \cite{direct_block_4}.} Accordingly, the AP possesses the reflective channels information, using them for optimally configuring the RIS for enhancing the communication performance. We exploit this reflective channel state information (CSI) to design the RIS parameters to make the aggregate reflective channels of the UTs to be as close as possible, making it suitable as a common source of randomness. The proposed algorithm further entails signal to noise ratio (SNR) maximization during the aggregate channel probing phase to mitigate the noise effect. With a linear probing complexity with respect to the number of users, our proposed method matches the efficiency of the broadcast-based methods. Moreover, we demonstrate that the proposed scheme achieves a high group KGR in quasi-static environments through a proper random phase shift setting in the RIS elements. This is a significant advantage over existing methods, which often struggle to attain high group KGR under static conditions \cite{JIoT_2022_Broadcast_Group}. The main contributions of this study are summarized as follows:
\begin{itemize}
\item For the first time, we incorporate an RIS to enable identical secret key generation in a group scenario. Leveraging the existing reflective CSI, we propose a novel GKG mechanism which offloads all computational tasks to the AP. This design ensures a lightweight solution for the GKG problem, making it particularly suitable for resource-constrained UTs. 
\item Our framework incorporates both PRIS and ARIS. While PRIS has been widely studied due to its low-power passive elements, we demonstrate that ARIS can more effectively align the aggregate channels of the UTs and mitigate additive white Gaussian noise (AWGN), owing to its element gains significantly greater than one.
\item To align the aggregate channels of the UTs while preserving the SNR, we propose an optimization method based on successive convex approximation (SCA) for the PRIS. For the ARIS, we adopt a novel semidefinite relaxation with Gaussian randomization (SDR-GR) approach.
\item We will observe that the probing complexity of our method scales linearly with the number of users making it suitable for relatively large networks. Moreover, after properly designing the RIS parameters, we can adjust the phase shifts and repeat the probings to further enhance the KGR in the environments with large coherence time. Unlike most of the existing studies, this capability ensures robust performance even in quasi-static channels.
\item We also provide a detailed discussion of the parameters influencing RIS-assisted GKG. Using various metrics such as normalized mean squared error (NMSE), key error rate (KER), KGR and key randomness we evaluate and benchmark the performance of the proposed method.
\end{itemize}     

The remainder of this paper is organized as follows. In Section \ref{sys_model}, the system model is defined. We review the pilot signal transmission for the key generation phase in Section \ref{pilot} and introduce the main parameters used in the paper. In Sections \ref{PRIS} and \ref{ARIS}, the proposed RIS-aided GKG method is expressed and the corresponding optimization problems are solved for the PRIS and ARIS, respectively. Section \ref{discussion}, provides further insights into the maximum number of supported UTs. Our numeric results are offered in \ref{numeric_discussion}, while our conclusions are presented in Section \ref{conclusion}.   
\section{System Model And Contributions} \label{sys_model}
\subsection{System Model And Approach}
We consider a group of $K$ single-antenna UTs aiming to generate identical secret key sequences based on their wireless channels, as illustrated in Fig. \ref{system_model}. A star topology is adopted, in which a central single-antenna AP, referred to as Alice, assists the UTs in generating identical keys with the aid of an RIS comprising $N$ elements. The RIS parameters are configured and controlled by Alice, and both Alice and the RIS are assumed to be trusted entities.

The primary challenge in group key generation arises from the inherent dissimilarity among the channels of different UTs. To address this issue, the proposed RIS-assisted approach consists of the following steps:

\subsubsection{Uplink Pilot Transmission}
We assume that the cascaded CSI corresponding to all RIS elements for each UT  is available at Alice. It is important to note that this CSI is not solely utilized for GKG and its primary role is to assist the UTs during the data transmission phase by enabling proper RIS configuration. The CSI is acquired via orthogonal pilot transmissions \cite{RIS_channel_estimation_TSP} from the UTs during the first phase, as illustrated in Fig. \ref{system_model}. Since the pilot sequences are assumed to be mutually orthogonal, all UTs can transmit their pilots within a single shared time interval \cite{RIS_channel_estimation_LWC}. For analytical tractability, we assume perfect CSI at Alice, while the impact of imperfect CSI on the proposed scheme is left for future investigation.

Numerous studies have addressed the problem of RIS channel estimation. One approach is to configure the phase shifts according to vectors of a fast Fourier transform (FFT) matrix, while transmitting pilot signals for each configuration. This approach requires $NQ$ timeslots $(Q\geq K)$ for channel estimation \cite{RIS_channel_estimation_TSP}, \cite{RIS_channel_estimation_LWC}. Alternatively, several works have proposed equipping the RIS with some RF chains to enable direct CSI acquisition. These methods require $Q+1$ timeslots $ (Q\geq K)$  for channel estimation \cite{ARIS_TWC_2021}, \cite{Alexandropoulos}, \cite{Alkhateeb}. As previously noted, the CSI acquisition phase for the RIS is primarily intended to support the data transmission phase and is not exclusive to the GKG process.

\subsubsection{Channel Alignment}
This is the core component of our novel GKG approach. Specifically, the main challenge in generating identical keys in a group setting stems from the dissimilarity among the channels between the UTs and AP. To address this issue, we exploit the subreflecting channel information obtained in the previous step to reduce the disparity among these channels as much as possible. Specifically, the RIS is leveraged to induce controllable modifications to the wireless propagation environment, thereby facilitating channel alignment across the UTs. The detailed design of the channel alignment strategy is presented in Sections \ref{PRIS} and \ref{ARIS} for the cases of passive PRIS and active ARIS, respectively.

\subsubsection{Downlink Pilot Transmission}
\begin{figure}
	\begin{center}
		\includegraphics[trim={3.5cm 0.5cm 4cm 1.3cm},clip,width=3.5in,height=2.4in]{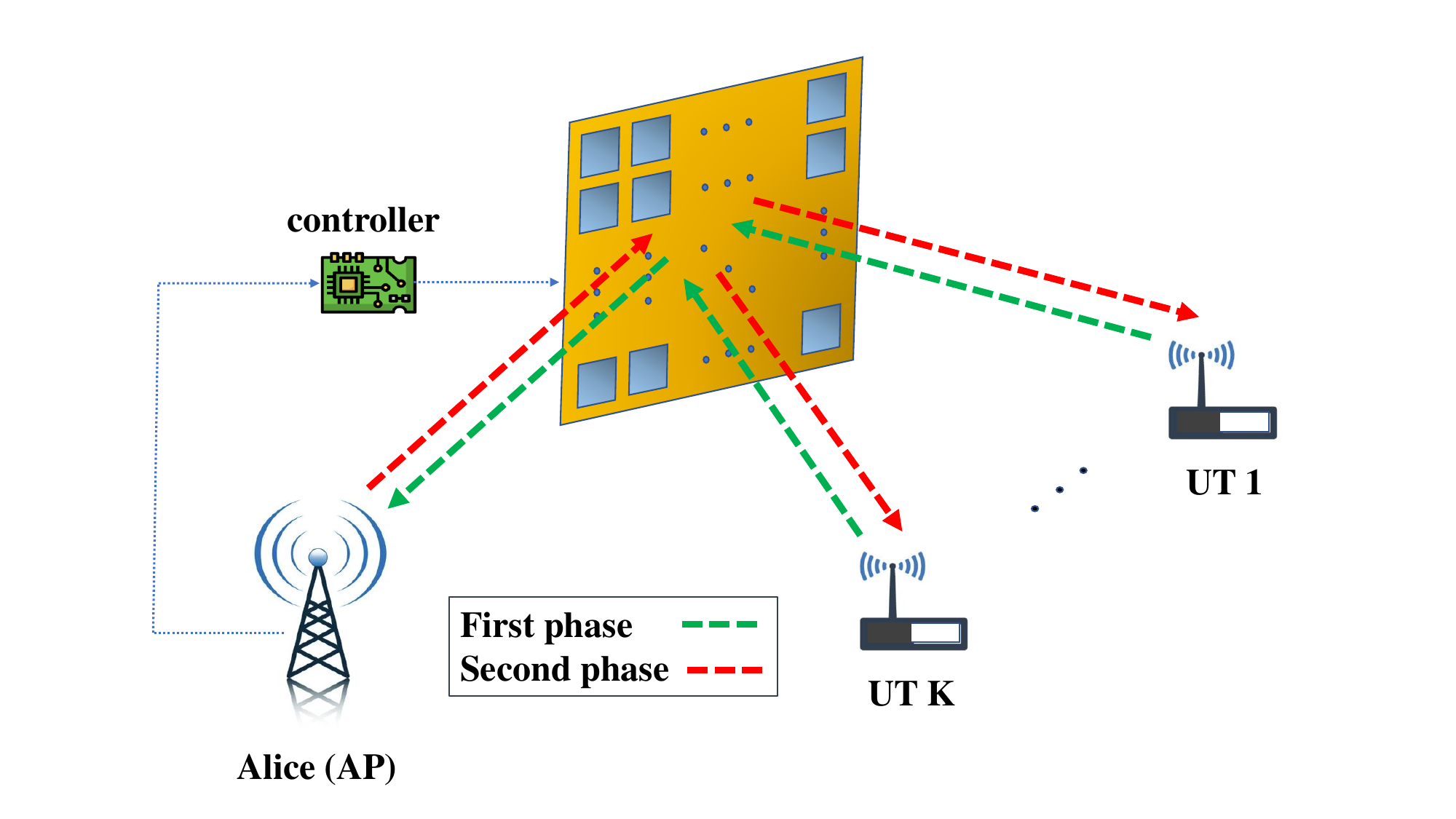}
		\caption{System model: Alice acting as an AP assists $K$ UTs with an RIS in obtaining identical group keys for secure communication.}
		\label{system_model}
	\end{center}
\end{figure}

After configuring the RIS parameters as described in the previous step, the aggregate reflected channels of the UTs become aligned. To establish a common source of randomness for group key generation, Alice broadcasts a pilot signal during the second phase, as illustrated in Fig.~\ref{system_model}. Each UT receives this signal after reflection from the RIS, and the resulting observations serve as the common randomness for secret key generation. The detailed procedures for this phase under both passive PRIS and ARIS configurations are presented in Section \ref{pilot}.
\subsection{Key Contributions and Advantages}   
\begin{figure}
	\begin{center}
		\includegraphics[trim={6.5cm 2.5cm 4.7cm 2.5cm},clip,width=3.7in,height=2.55in]{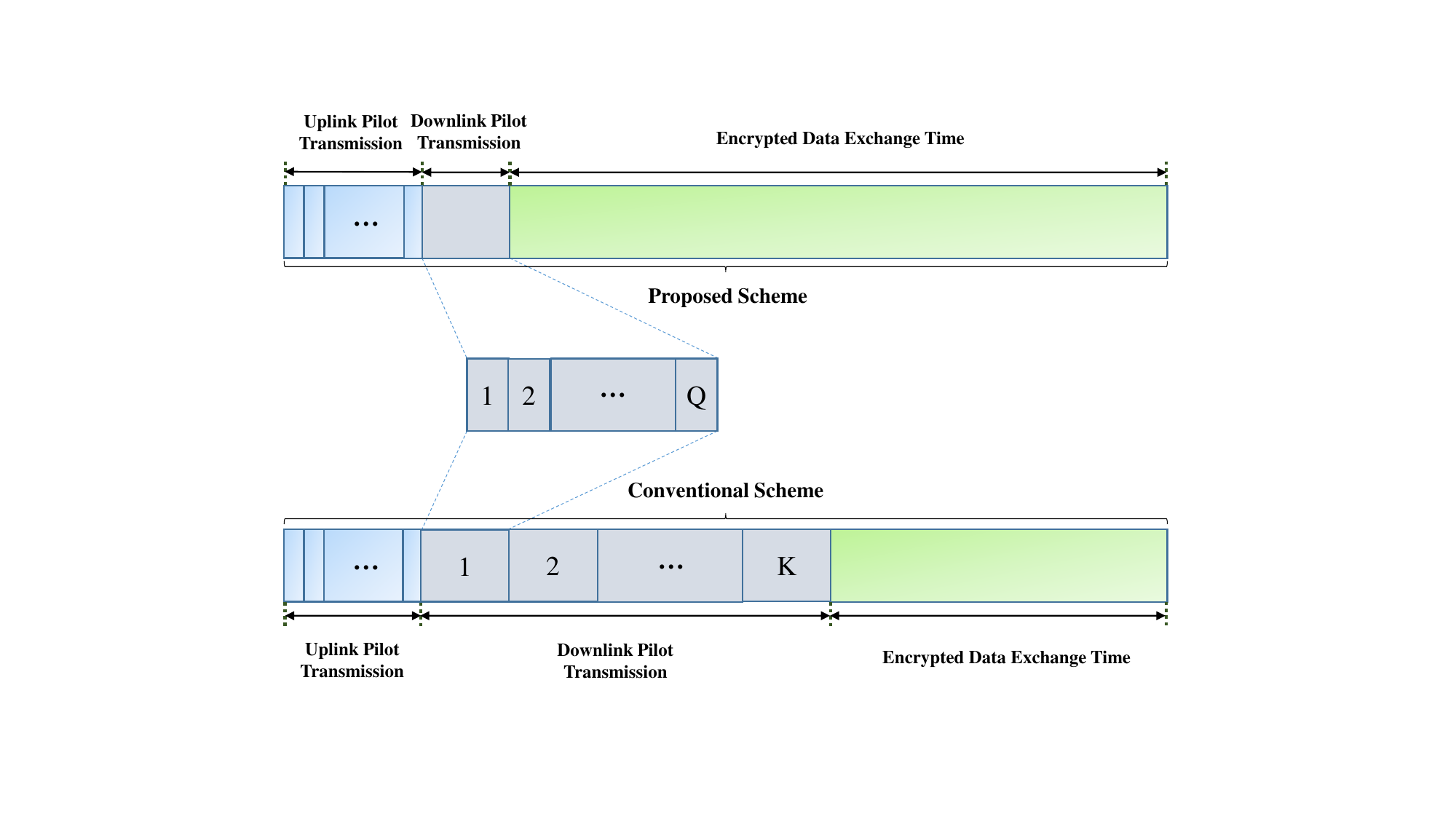}
		\caption{The proposed scheme against the conventional approach in \cite{LCOMM_2018_networkcoding}}
		\label{comparison}
	\end{center}
\end{figure}

Fig.  \ref{comparison} compares the efficiency of the proposed scheme with the conventional approach in \cite{LCOMM_2018_networkcoding}. In the conventional method, the UTs sequentially transmit pilot signals, allowing the AP to estimate their individual channels and establish $K$ pairwise keys, one with each UT. This leads to $K$ rounds of channel estimation which is so far equivalent to the first step of the proposed scheme. Then further steps are taken to agree on a common group key by exploiting the non-identical $K$ pairwise keys. As a result, a total of $2K$ channel estimation rounds are needed to generate the final common group key \cite{JIoT_2022_Broadcast_Group}, \cite{Zeng}. Moreover, in static environments, the achievable key rate of such approaches is typically limited due to the insufficient amount of common randomness that can be extracted from the wireless channel \cite{JIoT_2022_Broadcast_Group}.

In contrast, the proposed scheme eliminates the need for separate transmissions from the AP to each UT. Owing to the alignment of the channels between the AP and the UTs, a single round of pilot transmission is sufficient to provide all UTs with a common source of randomness for group key generation. Consequently, the proposed scheme requires only $K+1$ rounds of channel estimation, compared to $2K$ rounds in the conventional approach. Furthermore, channel alignment enables the AP to employ randomized pilot transmissions over multiple rounds to enhance the achievable key rate, particularly in static environments. This aspect will be discussed in more detail in the subsequent sections. It is important to emphasize that the CSI acquisition phase for the RIS elements is primarily intended to support the data transmission phase and is not exclusive to the GKG process.  Since it should not be regarded as an additional overhead of the proposed scheme. Nevertheless, even if the individual channel estimation phase is included as part of the GKG procedure, we saw that the required number of time slots scales linearly with the number of users, which is consistent with existing GKG methods \cite{JSYST_2022}. 
\section{Pilot Signal Transmission}\label{pilot}
Before proceeding to how to adjust the RIS parameters, we need to observe the pilot signal transmitted by Alice and received by the UTs in the second phase of transmission shown in Fig. \ref{system_model}. We consider both the PRIS and ARIS and will see their corresponding performance in the considered GKG scheme.
\subsection{Pilot Transmission In The PRIS Case}
The transmitted pilot by Alice is received in UT $k$ as
\begin{equation}\label{Alice_pilot_PRIS}
{\tilde{\mathbf{y}}_k}^{(P)}=\sqrt{P_a}\tilde{\mathbf{h}}_{kr}^H\mathbf{\Phi}\tilde{\mathbf{h}}_{ar}\mathbf{s}_a^H+\tilde{\mathbf{n}}_k,
\end{equation}
where $P_a$ is the transmit power of Alice, $\mathbf{\Phi}=\mathrm{diag}\left(\alpha_1e^{j\theta_1},\ldots,\alpha_Ne^{j\theta_N}\right)$ denotes the PRIS phase-shift matrix with continuous phase shifts $\{\theta_n\}$, and $\tilde{\mathbf{h}}_{ar}\sim\mathcal{CN}(\mathbf{0},\kappa_{ar}\mathbf{R})$ and $\tilde{\mathbf{h}}_{kr}\sim\mathcal{CN}(\mathbf{0},\kappa_{kr}\mathbf{R})$ are the channel vector coefficients from Alice and UT $i$ to the RIS, respectively. Additionally, $\kappa_{ar}=\sigma_{ar}^2d_Hd_V$ and $\kappa_{kr}=\sigma_{kr}^2d_Hd_V$ where $\sigma_{ar}^2$ and $\sigma_{kr}^2$ denote the large scale path loss between the RIS elements - Alice and $k$-th UT, respectively and $d_H$ and $d_V$ are the vertical height and horizontal width of each element. Moreover, $\mathbf{R}$ is covariance matrix between the RIS elements \cite{LWC_RIS_Bjornson}. Finally, $\mathbf{s}_a\in\mathbb{C}^{Q\times1}$ is the unit energy pilot signal of Alice with length $Q$ and $\tilde{\mathbf{n}}_k\sim\mathcal{CN}(\mathbf{0},\tilde{\sigma}_k^2\mathbf{I})$ is the AWGN at node $k$. To compensate for the large scale path loss difference between the nodes in the GKG process, we normalize $\tilde{\mathbf{y}}_k$ by $\kappa_{ar}$ and $\kappa_{kr}$ \cite{JIoT_2022_Broadcast_Group}. Accordingly, the estimated channel samples at the nodes is obtained as 
\begin{equation} \label{sample_Bob_PRIS}
y_k^{(P)}=\frac{{\tilde{\mathbf{y}}_k}^{(P)}\mathbf{s}_a}{\sqrt{P_a\kappa_{ar}\kappa_{kr}}}=\underbrace{\mathbf{h}_{kr}^H\mathbf{\Phi}\mathbf{h}_{ar}}_{h_k^{(P)}}+\underbrace{\frac{\tilde{n}_k}{\sqrt{P_aQ\kappa_{ar}\kappa_{kr}}}}_{n_k}.
\end{equation}
In \eqref{sample_Bob_PRIS}, $\mathbf{h}_{ar}\sim\mathcal{CN}(\mathbf{0},\mathbf{R})$, $\mathbf{h}_{kr}\sim\mathcal{CN}(\mathbf{0},\mathbf{R})$ and $n_k\sim\mathcal{CN}(0,\sigma_{n_k}^2=\tilde{\sigma}_k^2/(P_aQ\kappa_{ar}\kappa_{kr}))$. Notably, the independent channel vectors $\mathbf{h}_{kr}$ from each UT to the RIS is the main obstacle in maintaining the common randomness between the nodes for obtaining a group key sequence. Another setback is the effect of AWGN which hinders obtaining the group key by introducing channel estimation error (CEE)  regardless of managing to obtain the identical channel samples between the nodes.
\subsection{Pilot Transmission In The ARIS Case}
For the ARIS case, $k$-th UT obtains the pilot signal as
\begin{equation}\label{Alice_pilot_ARIS}
{\tilde{\mathbf{y}}_k}^{(A)}=\sqrt{P_a}\tilde{\mathbf{h}}_{kr}^H\mathbf{\Psi}\tilde{\mathbf{h}}_{ar}{\mathbf{s}_a}^H+\tilde{\mathbf{h}}_{kr}^H\mathbf{\Psi}\mathbf{N}_F+\tilde{\mathbf{n}}_k,
\end{equation}
where $\mathbf{N}_F\in\mathbb{C}^{N\times Q}$ is the amplification noise of the ARIS with zero mean and variance $\sigma_F^2$ and $\mathbf{\Psi}=\mathbf{\Lambda}\mathbf{\Theta}$ is the ARIS control matrix. $\mathbf{\Theta}=\mathrm{diag}(e^{j\theta_1},\ldots, e^{j\theta_n}, \ldots,e^{j\theta_N})$ is the phase shift matrix with continuous phases $\{\theta_n\}$, while $\mathbf{\Lambda}=\mathrm{diag}(\eta_1,\ldots, \eta_n,\ldots,\eta_N)$ denotes the amplification matrix. We note that the amplification factor for each element is $\eta_i=\sqrt{P_R^{(i)}/(P_a\lvert{\tilde{h}_{ar}}^{(i)}\rvert^2+\sigma_F^2)}$, ensuring that the output transmit power of each element is $P_R^{(i)}$. The total available transmit power of the ARIS is $P_R^{max}$ which satisfies $\sum\limits_{i=1}^{N}P_R^{(i)}\leq P_R^{max}$ \cite{ARIS_TWC_2021}. Moreover, due to the hardware limitations the amplification gain of each element is limited, namely, to $w_{max}$ \cite{ARIS_TWC_2021}. Finally, like the PRIS case we normalize the received signal by $\kappa_{ar}$ and $\kappa_{kr}$ and the channel samples at the nodes are obtained as
\begin{equation}\label{sample_Bob_ARIS}
y_k^{(A)}=\frac{{\tilde{\mathbf{y}}_k}^{(A)}\mathbf{s}_a}{\sqrt{P_a\kappa_{ar}\kappa_{kr}}}=\underbrace{\mathbf{h}_{kr}^H\mathbf{\Psi}\mathbf{h}_{ar}}_{h_k^{(A)}}+\underbrace{\frac{\tilde{\mathbf{h}}_{kr}^H\mathbf{\Psi}\mathbf{n}_F+\tilde{n}_k}{\sqrt{P_aQ\kappa_{ar}\kappa_{kr}}}}_{z_k},
\end{equation} 
where $z_k\sim\mathcal{CN}(0,\sigma_{z_k}^2=(\sigma_F^2\kappa_{kr}\mathrm{tr}(\mathbf{\Psi}\mathbf{\Psi}^H\mathbf{R})+\tilde{\sigma}_k^2)/(P_aQ\kappa_{ar}\kappa_{kr}))$. We further note that like the PRIS case, the major barrier in obtaining common randomness for generating group secret key is the independent channels of UTs to the ARIS, i.e., $\mathbf{h}_{kr}$. Similarly, the noise term present in the final channel sample, is another impediment in maintaining identical group keys.
\section{Acquiring Common Randomness With PRIS}\label{PRIS}
After acquiring the received pilot signals, in this section, we will discuss how we are going to exploit the ability of the PRIS in generating group keys. We will provide solutions to overcome the channel differences between the UTs within the group meanwhile mitigating the malicious effect of the AWGN. 

Firstly, we propose a metric to measure the channel difference between the UTs participating in the GKG process.  By letting
$\mathcal{K} = \{1,\ldots,K\}$,
the signal to error ratio (SER) between the channel sample of UT $k$ and the channel samples of the other UTs is defined as
\begin{align}\label{SER}
SER_{k,i}=\frac{\mathbf{v}^H\mathbf{\Sigma}_{ark}\mathbf{v}}{\lvert\mathbf{v}^H\mathbf{\Sigma}_{kr}\mathbf{h}_{ar}-\mathbf{v}^H\mathbf{\Sigma}_{ir}\mathbf{h}_{ar}\rvert^2},
\quad \forall k,i \in \mathcal{K},\; i \neq k
\end{align}
in which for the notational simplicity, we have defined $\mathbf{\Sigma}_{k(i)r}=\mathrm{diag}(\mathbf{h}_{k(i)r}^H)$, $\mathbf{\Sigma}_{ark}=\mathrm{diag}(\mathbf{h}_{kr}^H)\mathbf{h}_{ar}\mathbf{h}_{ar}^H\mathrm{diag}(\mathbf{h}_{kr})$.  $\mathbf{v}^H=[\phi_1,\ldots,\phi_N]$ and $\phi_i=\alpha_ie^{j\theta_i}$. According to \eqref{SER}, as the SER increases, the samples for GKG will become closer. We note that if the channel difference minimization was our only goal, we could set the reflecting coefficients equal to zero. However, applying this will merely result in the noise terms at the UTs and group key rate will inevitably be zero. Accordingly, we need to assure that the SNR at the UTs is considered as a part of our solution. Against this background, and by letting 
$\mathcal{N} = \{1,\ldots,N\}$,
we can write the corresponding optimization problem for the PRIS as
\begin{subequations}
\begin{align}
\hspace{-5mm}&\mathbf{P1:}\quad\max_{\mathbf{v}}\hspace{1mm}\min_{k\in\mathcal{K}} \quad \frac{\mathbf{v}^H\mathbf{\Sigma}_{ark}\mathbf{v}}{\sigma_{n_k}^2}\label{MDM_passive_obj}\\
&\textrm{s.t.}\quad  SER_{k,i}\geq SER_{th},\quad \forall k,i \in \mathcal{K},\; i \neq k ,\label{MDM_passive_cond1}\\
&\hspace{2.5mm}\mathbf{v}^H\mathbf{B}_i\mathbf{v}\leq1,\quad \forall i\in \mathcal{N},\label{MDM_passive_cond2}
\end{align}
\end{subequations}
where $SER_{th}$ is the minimum SER required to generate common secret keys and $\mathbf{B}_i=\mathbf{b}_i\mathbf{b}_i^H$ where $\mathbf{b}_i$ is a base vector with its $i$-th element being 1 and all the other elements being 0. In other words, \textbf{P1} tries to guarantee a certain level of SER for all the UTs while trying to boost the SNR for the worst UT to combat the AWGN. To solve \textbf{P1}, we can rewrite it as 
\begin{subequations}
\begin{align}
\hspace{-5mm}&\mathbf{P2:}\quad\max_{\mathbf{v},c} \quad c\label{MDM_passive_obj_2}\\
&\textrm{s.t.}\quad  c\leq \frac{\mathbf{v}^H\mathbf{\Sigma}_{ark}\mathbf{v}}{\sigma_{n_k}^2},\quad \forall k\in \mathcal{K}\label{MDM_passive_cond12}\\
&\hspace{2.5mm}\hspace{1mm}\mathbf{v}^H\mathbf{D}_{k,i}\mathbf{v}\leq\frac{\mathbf{v}^H\mathbf{\Sigma}_{ark}\mathbf{v}}{SER_{th}},\forall k,i \in \mathcal{K},\; i \neq k,\label{MDM_passive_cond22}\\
&\hspace{2.5mm}\mathbf{v}^H\mathbf{B}_i\mathbf{v}\leq1,\quad \forall i\in \mathcal{N},\label{MDM_passive_cond32}
\end{align}
\end{subequations}
where $\mathbf{D}_{k,i}=\left(\mathbf{\Sigma}_{kr}-\mathbf{\Sigma}_{ir}\right)\mathbf{h}_{ar}\mathbf{h}_{ar}^H\left(\mathbf{\Sigma}_{kr}-\mathbf{\Sigma}_{ir}\right)^H$. However, \eqref{MDM_passive_cond12} and \eqref{MDM_passive_cond22} are still non-convex. We can exploit
\begin{align}
\mathcal{F}_k^{(t)}(\mathbf{v})\delequal{\mathbf{v}^{(t)}}^H\mathbf{\Sigma}_{ark}\mathbf{v}^{(t)}+2{\mathbf{v}^{(t)}}^H\mathbf{\Sigma}_{ark}^H\left(\mathbf{v}-\mathbf{v}^{(t)}\right),
\end{align}
as the first order Taylor expansion of $\mathbf{v}^H\mathbf{\Sigma}_{ark}\mathbf{v}$ to convert the problem into a convex form. The final solution is given in Algorithm \ref{MDM}.
\begin{algorithm}
\caption{\hspace{-1mm}: Proposed optimization algorithm for the PRIS-aided GKG (\textbf{P2}) \label{MDM}}
\textbf{Input}: $\mathbf{\Sigma}_{kr}\mathbf{h}_{ar}$,\hspace{2mm}$k\in\mathcal{K}$.\\
\textbf{Output}: $\mathbf{v}$.
\begin{algorithmic}[1] 
\State \textbf{Initialization:}
\begin{itemize}
\item set the initial point as $\mathbf{v}^{(0)}=[0,\ldots,0]^T$
\end{itemize}
\Repeat
\State Update $\mathbf{v}^{(t)}$ with their values at step $t-1$ by solving:
\begin{subequations}
\begin{align}
\hspace{1mm}&\quad\max_{\mathbf{v},c} \quad c\label{MDM_passive_obj_3}\\
&\textrm{s.t.}\quad  c\leq \frac{\mathcal{F}_k^{(t)}(\mathbf{v})}{\sigma_{n_k}^2},\quad \forall k\in\mathcal{K}\label{MDM_passive_cond13}\\
&\hspace{2.5mm}\hspace{1mm}\mathbf{v}^H\mathbf{D}_{k,i}\mathbf{v}\leq\frac{\mathcal{F}_k^{(t)}(\mathbf{v})}{SER_{th}},\quad \forall k,i \in \mathcal{K},\; i \neq k,\label{MDM_passive_cond23}\\
&\hspace{2.5mm}\mathbf{v}^H\mathbf{B}_i\mathbf{v}\leq1,\quad \forall i\in \mathcal{N},\label{MDM_passive_cond33}
\end{align}
\end{subequations}
\State $t=t+1$
\Until the objective function converges.
\end{algorithmic}
\end{algorithm} 

Therefore, we have successfully convexified the optimization problem in \textbf{P1} with the SCA method. Note that the complexity order of the proposed SCA-based algorithm solved with MATLAB CVX is $\mathcal{O}(\log((K^2+N)/(\partial t^{(0)}))/\log{\epsilon})$ at each iteration of it, where $0\leq\partial\ll1$ is the stopping criterion for interior point method (IPM), $t^{(0)}$ is the initial point for the accuracy of approximation in IPM and $\epsilon$ is used for updating the accuracy of IPM \cite{Boyd,Mokari-1,Mokari-2}. 
\begin{remark}
We note that the individual CSI estimation is not originally a part of the proposed GKG process. This is because the primary goal of the RIS deployment is to assist the nodes in conveying their messages in the harsh wireless environment with no LoS paths. Achieving this goal requires individual CSI estimation for each element so that the AP is able to optimally design the phase shifts, for instance, to maximize sum throughput. Accordingly, this CSI is already available at the AP and we simply exploit it to generate the group key. For instance, the recent study in \cite{TIFs_Quasi_2026} assumes perfect CSI for the RIS elements to enhance the KGR while simultaneously satisfying the communication requirements; the RIS channel estimation process is not considered part of the KGR policy. Based on this assumption, only the timeslots allocated to the second phase of the transmission are specifically dedicated to the GKG. 
\end{remark}
\section{Acquiring Common Randomness With ARIS}\label{ARIS}
As the reflection factor of the ARIS elements is not limited to 1, we expect that the ARIS has a superior ability in compensating the channel differences between the UTs compared to PRIS. Moreover, due to its relaying capability, it can outperform the PRIS in combating the AWGN. This section deals with exploiting ARIS in our GKG scheme. Likewise, for the ARIS we have
\begin{subequations}
\begin{align}
\hspace{-5mm}&\mathbf{P3:}\quad\max_{\mathbf{w}}\hspace{1mm}\min_{k\in\mathcal{K}} \quad \frac{\mathbf{w}^H\mathbf{\Sigma}_{ark}\mathbf{w}}{\alpha_F\mathbf{w}^H\mathbf{\Sigma}_{kr}^H\mathbf{\Sigma}_{kr}\mathbf{w}+\alpha_{N_k}}\label{CDM_active_obj_4}\\
&\textrm{s.t.}\quad  SER_{k,i}\geq SER_{th},\quad \forall k,i \in \mathcal{K},\; i \neq k,\label{CDM_active_cond14}\\
&  P_a\mathbf{w}^H\mathbf{\Sigma}_{ar}^H\mathbf{\Sigma}_{ar}\mathbf{w}+\frac{\sigma_F^2\mathbf{w}^H\mathbf{w}}{\kappa_{ar}}\leq \frac{P_R^{max}}{\kappa_{ar}},\label{CDM_active_cond24}\\
&\hspace{2.5mm}\mathbf{w}^H\mathbf{B}_i\mathbf{w}\leq w_{max}^2,\quad \forall i\in \mathcal{N}.\label{CDM_active_cond34}
\end{align}
\end{subequations} 
where $\alpha_F=\sigma_F^2/(P_aQ\kappa_{ar})$, $\alpha_{N_k}=\tilde{\sigma}_k^2/(P_aQ\kappa_{ar}\kappa_{kr})$, $\mathbf{\Sigma}_{ar}=\mathrm{diag}(\mathbf{h}_{ar}^H)$, $\mathbf{w}^H=[\psi_1,\ldots,\psi_N]$ and $\psi_i=\eta_ie^{j\theta_i}$. Note that \eqref{CDM_active_cond24} ensures the total power consumption of the ARIS does not exceed $P_R^{max}$ while \eqref{CDM_active_cond34} limits the maximum gain of an ARIS element to $w_{max}$. To solve the GKG problem for the ARIS we can write \textbf{P3} as
\begin{subequations}
\begin{align}
\hspace{-5mm}&\mathbf{P4:}\quad\max_{\mathbf{w},c} \quad c\label{CDM_active_obj_5}\\
&\textrm{s.t.}\quad c\leq \frac{\mathbf{w}^H\mathbf{\Sigma}_{ark}\mathbf{w}}{\alpha_F\mathbf{w}^H\mathbf{\Sigma}_{kr}^H\mathbf{\Sigma}_{kr}\mathbf{w}+\alpha_{N_k}},\quad \forall k\in \mathcal{K},\\  
& SER_{k,i}\geq SER_{th},\quad \forall k,i \in \mathcal{K},\; i \neq k,\label{CDM_active_cond15}\\
&  P_a\mathbf{w}^H\mathbf{\Sigma}_{ar}^H\mathbf{\Sigma}_{ar}\mathbf{w}+\frac{\sigma_F^2\mathbf{w}^H\mathbf{w}}{\kappa_{ar}}\leq \frac{P_R^{max}}{\kappa_{ar}},\label{CDM_active_cond25}\\
&\hspace{2.5mm}\mathbf{w}^H\mathbf{B}_i\mathbf{w}\leq w_{max}^2,\quad \forall i\in \mathcal{N}.\label{CDM_active_cond35}
\end{align}
\end{subequations} 
By alternatively optimizing $\mathbf{w}$ and $c$, we can solve \textbf{P4}. The details of the proposed SDR-GR with alternate optimization (AO)  method is proposed in Algorithm \ref{ARIS-CDM}. 
\begin{algorithm}
\caption{\hspace{-1mm}: Proposed optimization algorithm for the ARIS-aided GKG (\textbf{P4}) \label{ARIS-CDM}}
\textbf{Input}: $c_{max}$, $\varepsilon$, $\mathbf{\Sigma}_{kr}, \mathbf{h}_{ar}$,  $k\in\mathcal{K}$.\\
\textbf{Output}: $\mathbf{w}$.
\begin{algorithmic}[1] 
\State \textbf{Initialize} $c_{min}=0$. 
\Repeat{ (Bisection search for $c$)}
\State set $c^{(t)}=\left(c_{max}+c_{min}\right)/2$.
\State \textbf{Find} the feasible ARIS beamforming vector $\mathbf{w}$ . 
\State For given $c^{(t)}$, find the solution for the below relaxed feasibility problem :
\begin{subequations}
\begin{align}
\hspace{-5mm}&\quad\max_{\mathbf{W},c} \quad c\label{CDM_active_obj_6}\\
&\textrm{s.t.}\quad c\left(\alpha_F\mathrm{Tr}\left\{\mathbf{\Sigma}_{kr}^H\mathbf{\Sigma}_{kr}\mathbf{W}\right\}+\alpha_{N_k}\right)\leq \nonumber\\
&\hspace{20mm}\mathrm{Tr}\left\{\mathbf{\Sigma}_{ark}\mathbf{W}\right\},\quad \forall k\in \mathcal{K},\label{CDM_active_cond16}\\  
& \mathrm{Tr}\left\{\mathbf{D}_{k,i}\mathbf{W}\right\}\leq\frac{\mathrm{Tr}\left\{\mathbf{\Sigma}_{ark}\mathbf{W}\right\}}{SER_{th}},\quad \forall k,i \in \mathcal{K},\; i \neq k,\label{CDM_active_cond46}\\
&  P_a\mathrm{Tr}\left\{\mathbf{\Sigma}_{ar}^H\mathbf{\Sigma}_{ar}\mathbf{W}\right\}+\frac{\sigma_F^2\mathrm{Tr}\left\{\mathbf{W}\right\}}{\kappa_{ar}}\leq \frac{P_R^{max}}{\kappa_{ar}},\label{CDM_active_cond56}\\
&\hspace{5mm}\big\lvert\mathrm{Tr}\left\{\mathbf{B}_i\mathbf{W}\right\}\big\rvert\leq w_{max}^2,\quad \forall i\in\mathcal{N},\label{EEM_active_cond66}\\
&\hspace{5mm}\mathbf{W}\succeq \mathbf{0}\label{EEM_active_cond76}
\end{align}
\end{subequations} 
\State Initiate the GR and find the rank one solution.
\If{The SDP problem is solvable}{ record $\mathbf{w}_{opt}=\mathbf{w}$ and set $c_{min}=c^{(t)}$.}
\Else{ $c^{(t)}$ is unreachable and set $c_{max}=c^{(t)}$.}
\EndIf
\State $t=t+1$
\Until The difference $(c_{max}-c_{min})$ is less than $\varepsilon$. 
\end{algorithmic}
\end{algorithm} 

In Algorithm \ref{ARIS-CDM}, by fixing the value of $c$ and setting $\mathbf{W} = \mathbf{w}\mathbf{w}^H$, \textbf{P4} converts to a feasibility problem with semidefinite programming (SDP) form as the rank one constraint $\mathrm{rank}(\mathbf{W})=1$ is missing. Using MATLAB CVX we can obtain the relaxed solution, namely, $\tilde{\mathbf{W}}$. Then we can use the GR to obtain a suboptimal rank-one solution. Specifically, by performing the eigenvalue decomposition we obtain $\tilde{\mathbf{W}}=\mathbf{U}\mathbf{\Sigma}\mathbf{U}^H$ where the columns of $\mathbf{U}$ are the eigenvectors of $\tilde{\mathbf{W}}$ and $\mathbf{\Sigma}$ is a diagonal matrix including the eigenvalues of $\tilde{\mathbf{W}}$. We generate 1000 instances of random candidate vectors, i.e., $\tilde{\mathbf{w}}_i=w_{max}*\frac{\mathbf{z}_i}{\mathrm{max}\left\{\lvert\mathbf{z}_i\rvert\right\}}$, $\mathbf{z}_i=\mathbf{U}\mathbf{\Sigma}^{\frac{1}{2}}\boldsymbol{\zeta}_i$, $i = 1,\ldots,1000$ where $\boldsymbol{\zeta}_i\sim\mathcal{CN}(\mathbf{0},\mathbf{I}_N)$. Then, $\tilde{\mathbf{W}}_i=\tilde{\mathbf{w}}_i\tilde{\mathbf{w}}_i^H$ which satisfies \eqref{CDM_active_cond16}, \eqref{CDM_active_cond46}, \eqref{CDM_active_cond56} and results in the maximum of minimum SNR among the users is chosen as the SDP output. Note that to guarantee the convergence of the AO it is essential to perform a large number of Gaussian randomizations \cite{LWC_2023_GR}. This ensures the space is adequately searched to find the solution to the feasibility problem. We further note that our proposed SDR-GR method can solve the ARIS problem with the worst-case complexity of $\mathcal{O}((K^2+N+1)^4N^{1/2}\log(1/\varepsilon))$ \cite{MSP_2010_SDR} at each iteration of the algorithm. We have employed Sedumi solver for the SDR problem in our numerical results because of its superior performance in our problem. This solver utilizes matrix sparsity yielding lower complexity than the stated result \cite{MSP_2010_SDR}.

As mentioned earlier, the ARIS has higher ability in reaching identical channel samples for group PLKG than the PRIS as it can reach the element gains greater than 1 thanks to its active elements. Additionally, it can fulfill the SNR requirement more easily than the PRIS for the same reason.
\begin{remark}
We note that the arithmetic complexity of the two algorithms for the PRIS and ARIS should not be mixed up with the \textbf{probing complexity} of the group key algorithm. The probing complexity \cite{TVT_2019_OFDMA} of our proposed group key agreement is $\mathcal{O}(Q), Q\geq K$ which grows linearly with the number of users if we consider the individual channel probing as a part of it. To elaborate further, the computational complexity is transferred to the AP and the resource-limited UTs are not involved in the process of designing the beamforming vector for the common key generation.
\end{remark}
\begin{remark}
 We demonstrate that the proposed scheme achieves a high group KGR in quasi-static environments by inducing randomness through either Alice or an appropriate random phase shift configuration at the RIS elements. This constitutes a significant advantage over existing approaches, which often struggle to attain high group KGR under static conditions \cite{JIoT_2022_Broadcast_Group}. Specifically, after configuring the RIS parameters using the corresponding algorithms, the channels to all UTs are equalized. Subsequently, Alice can introduce a random phase shift to its pilot signals, or the RIS can apply an identical random phase shift across all its elements. This operation results in a new phase offset when combined with the phase of the previously aligned channel samples, thereby creating a new source of common randomness that can be used to generate identical group keys. Notably, the samples observed by the eavesdropper (Eve) are not aligned with those of the UTs, since Eve's channel is not equalized with theirs. Consequently, Eve cannot obtain any useful information about the newly generated channel samples \cite{Aldaghri}.  
\end{remark}
\section{Discussion On The Number Of Users}\label{discussion}
We'll now further investigate the boundaries of our proposed GKG scheme concerning the number of users. Accordingly, let's consider the high available power regime where we aim to enforce equality among the aggregate reflected channels of the UTs. In other words 
\begin{equation}
\begin{aligned}
\mathbf{x}^H\mathbf{\Sigma}_{1r}\mathbf{h}_{ar} &= c, \\
\vdots \hspace{11mm}&\hspace{5mm} \vdots \\
\mathbf{x}^H\mathbf{\Sigma}_{Kr}\mathbf{h}_{ar} &= c,
\end{aligned}
\label{linear_system}
\end{equation}
where $\mathbf{x}=\mathbf{v}$ for the PRIS case, $\mathbf{x}=\mathbf{w}$ for the ARIS case and $\lvert c\rvert$ is sufficiently large to achieve a desired KER. The set of equations in \eqref{linear_system} represents a linear system with $K$ equations and $N$ variables. In the high available power regime, where $w_{max}\gg1$, \eqref{CDM_active_cond24} and \eqref{CDM_active_cond34} become irrelevant for the ARIS, meaning there are no constraints on the variables of the system. This implies that the system can be solved as long as $K\leq M$. However, for the PRIS, the inherent hardware limitation described in \eqref{MDM_passive_cond2} still exists, preventing the system from being solved for large $K$, i.e., $K=M$. In both scenarios, if $K>M$, a sufficient SNR cannot be procured, leading to poor system performance. These boundaries will be further reported in our numerical results. 
\section{Numerical Results And Discussions}\label{numeric_discussion}
In this section, we discuss the performance of the proposed RIS-aided GKG scheme. First, we introduce the performance metrics. Then, we present and analyze the numerical results.
\subsection{Performance Metrics}
\subsubsection{Normalized Mean Square Error}
To evaluate the effectiveness of the proposed method in aligning the received samples of users participating in GKG and its resilience to AWGN, the NMSE metric is employed. This metric is widely used in the literature to assess the efficacy of common randomness enforcement techniques \cite{FDD_JIoT_Guyue}, \cite{FDD_TVT_Guyue}. Accordingly, we consider sufficiently large instances of channels and for each instance we apply our scheme. The normalized squared error (NSE) between each user pair is computed and the maximum NSE is selected to account for the worst-case misalignment. The NMSE is then obtained by averaging these maximum NSE values across all instances. In other words we have
\begin{align}
\mathrm{NSE}_{i,j}=\frac{\lvert y_i^{(X)}-y_j^{(X)}\rvert^2}{\lvert y_i^{(X)}\rvert^2}, \quad & i,j\in\mathcal{K}, i\neq j\nonumber\\&, X\in\{P,A\},
\end{align}
and
\begin{equation}
\mathrm{NMSE} = \mathbb{E}\left[\max\left({\mathrm{NSE}_{i,j}}\right)\right].
\end{equation}
\subsubsection{Key Error Rate (KER)}
\begin{figure}
	\begin{center}
		\includegraphics[trim={6cm 4.5cm 4.7cm 0.5cm},clip,width=3.7in,height=2.55in]{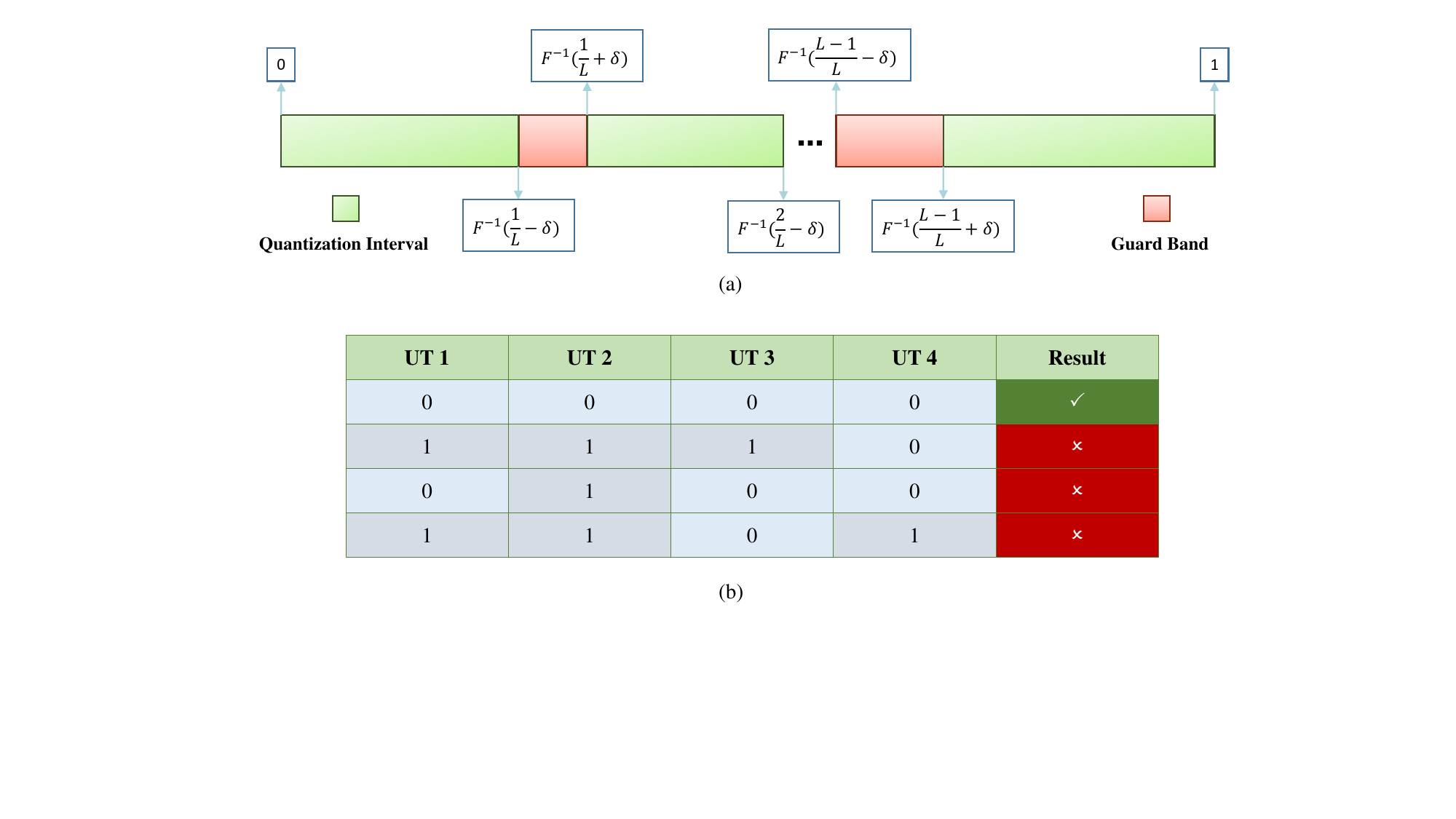}
		\caption{(a) Quantization intervals and the corresponding guard bands. (b) KER calculation in the group scenario.}
		\label{ker-quant}
	\end{center}
\end{figure}

To further investigate the KER of the proposed method, we have quantized the aligned channel samples based on the approach in \cite{FDD_JIoT_Guyue}. This method is based on the equal probability quantization and it entails quantization of both amplitude and phase information of the channel samples. Moreover, it provides guard band to control the KER. Specifically, we form the channel features by stacking real and imaginary parts of the aligned samples and normalize them. Now we are dealing with real numbers in the interval $[0,1]$. The mean and variance of the features are calculated and without loss of generality, it is assumed they have Gaussian distribution. If we consider $L$ levels of quantization, the $l$-th quantization interval is calculated as
\begin{equation}
[F^{-1}(\frac{l-1}{L}+\delta),F^{-1}(\frac{l}{L}-\delta)],\quad l= 2,\ldots,L-1,
\end{equation}
while the first quantization interval is $[0,F^{-1}(\frac{1}{L}-\delta)]$ and $L$-th quantization interval is $[F^{-1}(\frac{L-1}{L}+\delta),1]$. Fig. \ref{ker-quant}(a) shows a sketch of this approach. Note that $F^{-1}$ is the inverse of the cumulative distribution function (CDF) of the channel features, $\delta\in(0,1/2L)$ is the quantization factor used to set the limit of the guard band.  Finally, we use the common binary encoding to convert the channel features into a bit stream.

Now that the channel samples have been converted into binary sequences, an error in the group scenario is defined as the occurrence of at least one inconsistent bit among the UTs. Conversely, a correct case is one in which all bits are identical across the UTs. This is visually illustrated in Fig. \ref{ker-quant}(b). The KER is defined as the ratio of the total number of erroneous bits to the total number of key bits. Note that samples falling within the guard bands are discarded and are not considered in the KER calculation. Selecting wider guard bands represents a conservative approach, which can improve KER performance. The impact of this approach on the KGR will be analyzed subsequently.
\subsubsection{Key Generation Rate (KGR)}
To analytically evaluate the KGR of the proposed scheme, it is necessary to characterize the statistical distribution of the generated channel samples. However, since the RIS is configured based on the instantaneous CSI of the UTs, deriving a tractable analytical expression for the distribution of the resulting samples used in common key generation is not feasible. Similar challenges have been reported in prior studies. Therefore, in this work, we adopt the numerical approach for KGR evaluation proposed in \cite{FDD_JIoT_Guyue}, \cite{FDD_TVT_Guyue}. We first review the methodology and then discuss the challenges associated with its extension to the group scenario.

As noted earlier, samples falling within the guard bands are discarded, implying that no key bits are extracted from the corresponding probing attempts. The remaining samples, which lie within the quantization intervals, are mapped to bits, and any inconsistencies among these bits can be resolved during the reconciliation phase, provided that the KER remains below a specified target \cite{FDD_JIoT_Guyue}. Accordingly, the KGR is defined as the ratio of the initially extracted key bits to the total number of key generation attempts, subject to satisfying the target KER constraint.

Assume that an error detection protocol–based approach (EDPA) \cite{frontier_Access_2020} is employed for information reconciliation in the group scenario. Specifically, one of the UTs is selected to generate parity bits based on its obtained bit stream and broadcast them to the other UTs. Let $P_b$ denote the bit mismatch probability between the selected UT and any other UT. Accordingly, defining the group error probability as $P_g$, we can express it as follows:
\begin{equation}\label{P_g}
P_g = 1-(1-P_b)^{(K-1)},
\end{equation}
This expression for $P_g$ follows directly from the KER definition presented in the previous part. Through straightforward mathematical manipulation, it can be further expressed as follows:
\begin{equation}\label{P_b}
P_b = 1-e^{\frac{\ln\left(1-P_g\right)}{K-1}}.
\end{equation}
By a recursive argument, it can be shown that $P_b\leq P_g$. Accordingly, since the target KER for the two-user scenario in \cite{FDD_JIoT_Guyue} is applied to the proposed group scheme in the numerical results ($P_g\leq P_b^t$), the UTs are able to resolve their inconsistencies.
\subsubsection{Randomness}
We will investigate the various metrics to check for the randomness of the generated bit streams. We will exploit the National Institute of Standards and Technology (NIST) statistical test suite \cite{NIST} for the randomness tests. 
\subsection{Numerical Results}
In this subsection, we discuss the crucial parameters affecting our GKG scheme and present the corresponding numerical results. In the following discussions, we assume square-shaped RIS with the element sizes $d_H=d_V=\lambda/2$ where $\lambda$ is the carrier wavelength \cite{LWC_RIS_Bjornson}.  We further assume an indoor propagation environment. As stated in Section \ref{pilot}, the CSI measurements of each UT are normalized by its respective large scale path loss coefficient to compensate for differences in distances of nodes from Alice. Accordingly,  to ensure a fair comparison across scenarios having different number of UTs, we set $d_{kr}=70$ m and $d_{ar} = 50$ m. We also assume that the nodes are sufficiently separated such that their channels can be considered statistically independent. This corresponds to a worst-case scenario, where the RIS faces challenges in aligning the aggregate channels of the UTs due to their complete independence. Furthermore, the channel variances are given by $\sigma_{ir}^2=G_i-10\zeta_{ir}\log_{10}(d_{ir}/d_0)+\sigma_0^2$, where $i,j\in\left\{a,k\right\}$, $G_i=4$ dBi denotes the antenna gain at Alice and Bob, $\sigma_0^2=-30$ dB is the path loss at $d_0=1$ m and $\zeta_{ir}=2.2$ is the path loss exponent \cite{JIoT-2023-Vahid}. Moreover, we set the carrier frequency as $f_c = 1$ GHz, the bandwidth BW = 100 KHz, the noise figure $\mathrm{NF}_k = 5$ dB , $\sigma_F^2=\tilde{\sigma}_k^2$, $w_{max}^2 = 40$ dB and $Q=20$ \cite{ARIS_TWC_2021}. 

\begin{figure}
	\begin{center}
		\includegraphics[width=\linewidth, keepaspectratio]{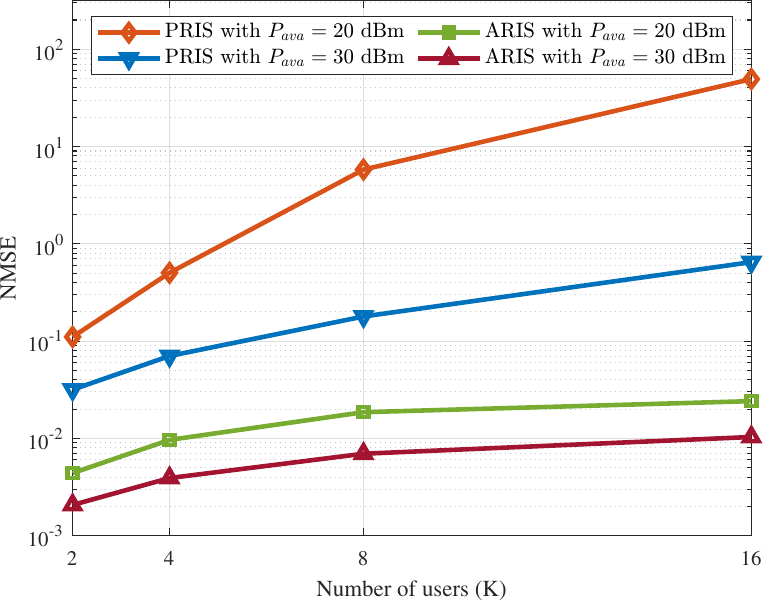}
		\caption{NMSE of the proposed GKG method versus group size for $N=64$ and $SER_{th} = 15$ dB.}
		\label{NMSE_vs_K}
	\end{center}
\end{figure} 
Fig. \ref{NMSE_vs_K} illustrates the NMSE performance of the proposed GKG method for both PRIS and ARIS under two available power levels. To ensure a fair comparison, the total available power ($P_{ava}$) is kept identical for both scenarios that is in the ARIS case, the available power is split equally between the AP and the RIS ($P_a=P_R^{max}=P_{ava}/2$) while for the PRIS $P_a=P_{ava}$. We can observe that the ARIS significantly outperforms PRIS in aligning user channel samples and combating AWGN. We further observe that as the number of users increases, the alignment capability degrades due to the fixed number of RIS elements, which limits the effectiveness of Alg. \ref{MDM} and Alg. \ref{ARIS-CDM}. Furthermore, the PRIS exhibits greater sensitivity to transmit power variations compared to the ARIS, as its performance improves more substantially with increased power. This suggests that the PRIS is more susceptible to AWGN owing to its double-fading effect.  

\begin{figure}
	\begin{center}
		\includegraphics[width=\linewidth, keepaspectratio]{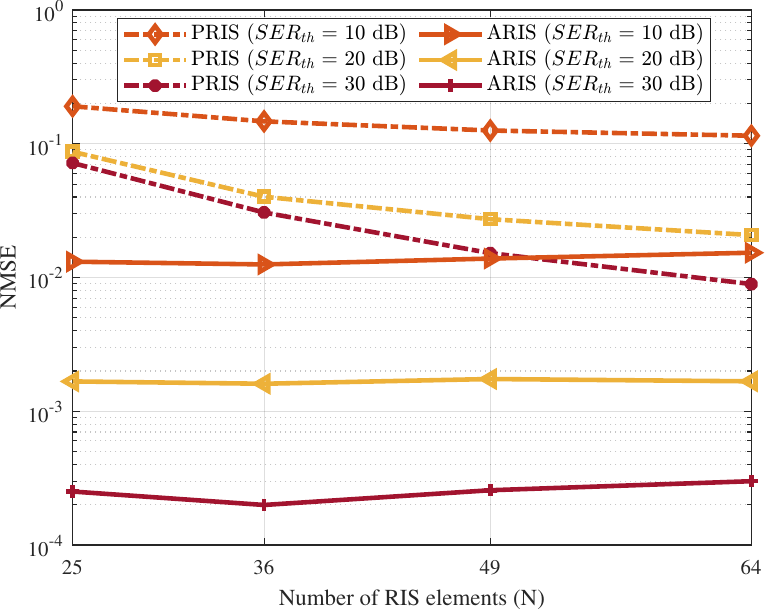}
		\caption{NMSE of the proposed GKG method versus number of RIS elements for $K=4$ and $P_{ava}=36$ dBm.}
		\label{NMSE_vs_N}
	\end{center}
\end{figure} 
Fig. \ref{NMSE_vs_N}, shows the NMSE performance of the proposed method versus number of RIS elements for different $SER_{th}$ values. As expected, the NMSE trend is non-increasing as adding the elements to the RIS will enhance both the SNR and sample alignment capability of the RIS. Note that as the amplification gain of the ARIS is not limited to one and it is capable of combating the double-fading effect, it manages to reach the target level of corresponding $SER_{th}$ in high SNR for $N=25$. Accordingly, increasing the number of elements has marginal effect on decreasing the NMSE as the system is already in the high SNR regime. This shows that $SER_{th}$ is the bottleneck in reaching identical channel samples for common randomness. This is shown in the substantial decrease in the NMSE value when the $SER_{th}$ increases. It is intuitive that imposing the extra condition on the samples to become close to each other will result in the SNR decline. However, for the ARIS or the PRIS with large number of elements that can acquire high SNR, having high $SER_{th}$ levels can result in better system performance. Our numerical results show that the decline in SNR level by increasing the $SER_{th}$ is comparably low when the system has high degrees of freedom (typically $K\leq N/4$ for PRIS). Accordingly, as the combined effect of SNR and SER determines the final system performance, it is generally recommended to consider high $SER_{th}$ in the proposed GKG method.          

\begin{figure}
	\begin{center}
		\includegraphics[width=\linewidth, keepaspectratio]{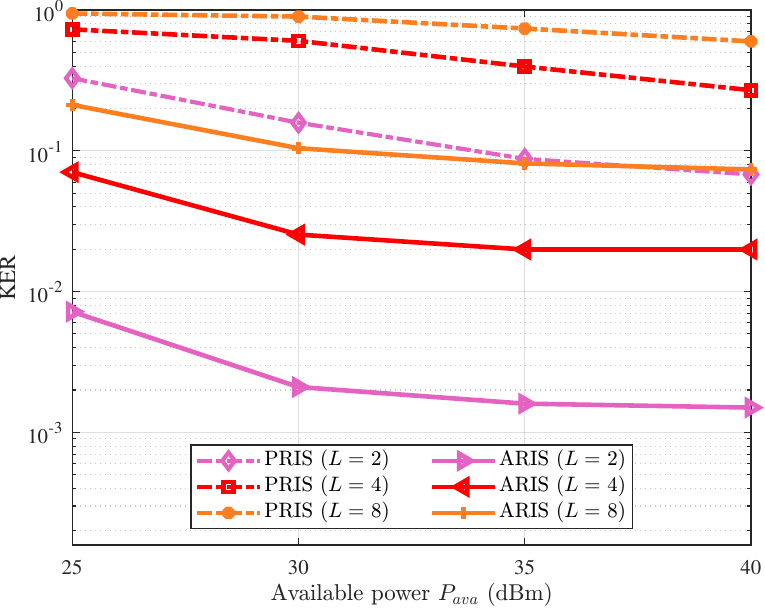}
		\caption{KER of the proposed GKG method versus available power for $N=16$, $K=4$, $\nu=0.02$ and $SER_{th} = 15$ dB.}
		\label{KER_vs_P}
	\end{center}
\end{figure} 
Based on the previously mentioned quantization method, the KER performance of our GKG scheme is shown in Fig. \ref{KER_vs_P}. This figure shows the performance of the initial key for three levels of quantization, namely, $L=2,4,8$. The value of the guard band is set as $\delta=\frac{0.02}{2(L-1)}$. It can be observed that in all of the three quantization levels the KER performance of ARIS is better than $10^{-1}$ while the PRIS showcases this KER performance only for $L=2$ at high transmit powers. Notably, the commonly  used information reconciliation methods can correct the initial key with KER less than $P_b^t = 10^{-1}$ \cite{FDD_JIoT_Guyue}. According to Fig. \ref{KER_vs_P}, for an equal total power budget, the ARIS leads to significantly lower KER than the PRIS.   

\begin{figure}
	\begin{center}
		\includegraphics[width=\linewidth, keepaspectratio]{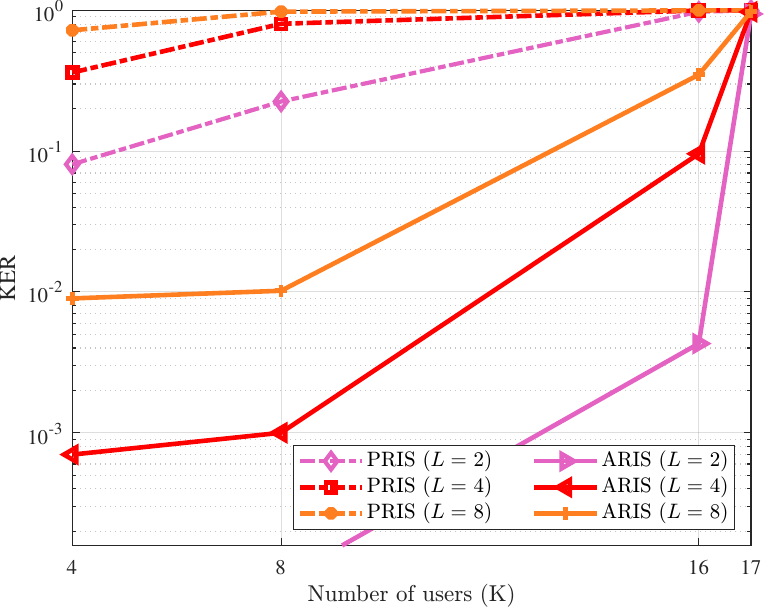}
		\caption{KER of the proposed GKG method versus number of the UTs for $N=16$, $P_{ave}=35$ dB, $\nu=0.02$ and $SER_{th} = 25$ dB.}
		\label{KER_vs_K}
	\end{center}
\end{figure} 
The KER versus the number of UTs within the group is illustrated in Fig. \ref{KER_vs_K}. While the PRIS achieves KER$\leq 10^{-1}$ only for $L=2$ when the number of users is $K=4$, the ARIS demonstrates significantly better performance at higher $L$, even when $N=K=16$. This observation is consistent with the discussion in Section \ref{discussion}. The ARIS can enforce channel sample equality among UTs while preserving SNR, due to its amplitude gain being significantly greater than one. In contrast, the intrinsic hardware limitations of PRIS prevent it from achieving this capability as the number of users approaches $N$. For both cases, the KER increases substantially when $N>K$, indicating that the maximum number of UTs the proposed scheme can support within a group is $K=N$. 

\begin{figure}
	\begin{center}
		\includegraphics[width=\linewidth, keepaspectratio]{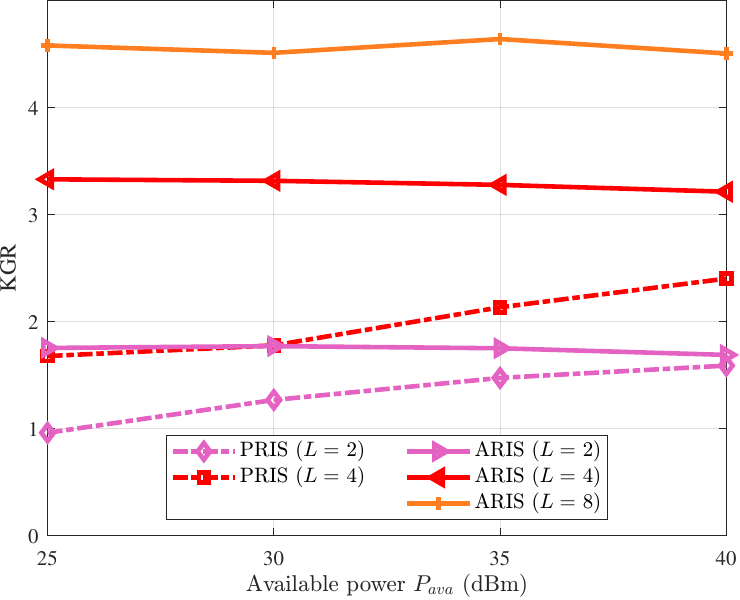}
		\caption{KGR of the proposed GKG method versus available power for $N=16$, $K=4$, $\nu=0.2$ and $SER_{th} = 25$ dB.}
		\label{KGR_vs_P}
	\end{center}
\end{figure} 
\begin{table}
\centering
\caption{Minimum available power required to reach KER = $10^{-1}$ with $\nu=0.2$ ($N=16$, $K=4$, $SER_{th}=25$ dB)}
\label{KER_vs_P_table}
\begin{tabular}{|c|c|c|c|}
\hline
\textbf{Quantization Level} & $L=2$ & $L=4$ & $L=8$ \\ 
\hline \hline
\textbf{PRIS (dBm)} &  21  & 37 & -\\ 
\hline 
\textbf{ARIS (dBm)} &  13 &  17 &  22\\ 
\hline
\end{tabular}
\end{table}  
Fig. \ref{KGR_vs_P} shows the KGR performance of our GKG method. $\delta=\frac{\nu}{2(L-1)}$ means that regardless of the quantization level, $\nu$ of the channel characteristics are discarded. In our results, we choose $\nu=0.2$ which means $20\%$ of the channel features are deleted during the quantization process. This leads the number of quantized bits generated reduce from $\log_2L$ to $0.8\log_2L$. Table \ref{KER_vs_P_table} shows the minimum available power required to reach KER = $10^{-1}$ for the ARIS and PRIS. According to Fig. \ref{KGR_vs_P}, in the PRIS case, when the available power is low, a considerable portion of features will be in the guard band resulting in lower KGR. However, when the transmit power increases, the algorithm performance enhances and the KGR is close to its maximum for $L=2$. Notably, the ARIS has a relatively high SNR due its resilience to double-fading effect and increasing the transmit power does not impact its KGR dramatically. The fluctuations associated with both cases are marginal and the KGR of PRIS for $L=2$ and ARIS for all three cases of $L$ is close to its maximum. This shows that the Gaussian distribution is a proper fit to the aligned channel features \cite{FDD_JIoT_Guyue}, \cite{FDD_TVT_Guyue}. Note that for an equal transmit power budget, the ARIS can lead to four times higher group KGR than the PRIS.

\begin{table}
\centering
\caption{Pass ratio test results of NIST random test suite ($N=16$, $K=4$, $P_{ave}=35$ dB).}
\label{NIST}
\begin{tabular}{|c|c|c|}
\hline
\textbf{Test} & \textbf{PRIS} & \textbf{ARIS} \\ 
\hline \hline
Approximate Entropy &  0.88 & 0.9\\ 
\hline 
Block Frequency & 0.93  & 0.97 \\ 
\hline
Cumulative Sums &  0.77 & 1.00 \\
\hline
FFT &  0.99 & 0.99 \\
\hline
Frequency & 0.74  & 1.00 \\
\hline
Ranking &  0.99 & 0.99 \\
\hline
Runs &  0.95 &  1.00\\
\hline
Longest Run & 0.92  &  0.99 \\
\hline
Serial & 0.96  & 0.99\\
\hline
\end{tabular}
\end{table}  
The NIST statistical test suite is widely used to evaluate the randomness of bit sequences, including those generated through PLKG \cite{FDD_JIoT_Guyue}, \cite{FDD_TVT_Guyue}. Here, we assess the randomness of the generated key sequences by performing nine tests from the suite. Although the NIST suite consists of 15 tests, some of them require very long input sequences e.g., $10^6$ bits. Obtaining such  long sequences is often challenging in both simulations and experimental settings. Therefore, only a subset of the tests is typically employed to evaluate the statistical properties of the generated sequences \cite{frontier_Access_2020}. Since generating long sequences is computationally expensive in our scheme as well, we consider nine tests in this study. Specifically , the entire sequence is first divided into multiple smaller bitstreams, each of which is subjected to all nine tests. This yields a $p$-value, and a bitstream is considered to pass a test if its $p$-value exceeds $0.01$. The pass ratio is defined as the number of bitstreams that pass a given test divided by the total number of bitstreams. Table \ref{NIST} presents the results of the NIST test suite applied to $105$ bitstreams, each consisting of $2000$ bits, generated by Algorithms \ref{MDM} and \ref{ARIS-CDM}. It can be observed that, for most of the tests, a significant number of bitstreams pass the $p$-value criteria.

\begin{figure*}
	\begin{center}
		\includegraphics[width=\linewidth, keepaspectratio]{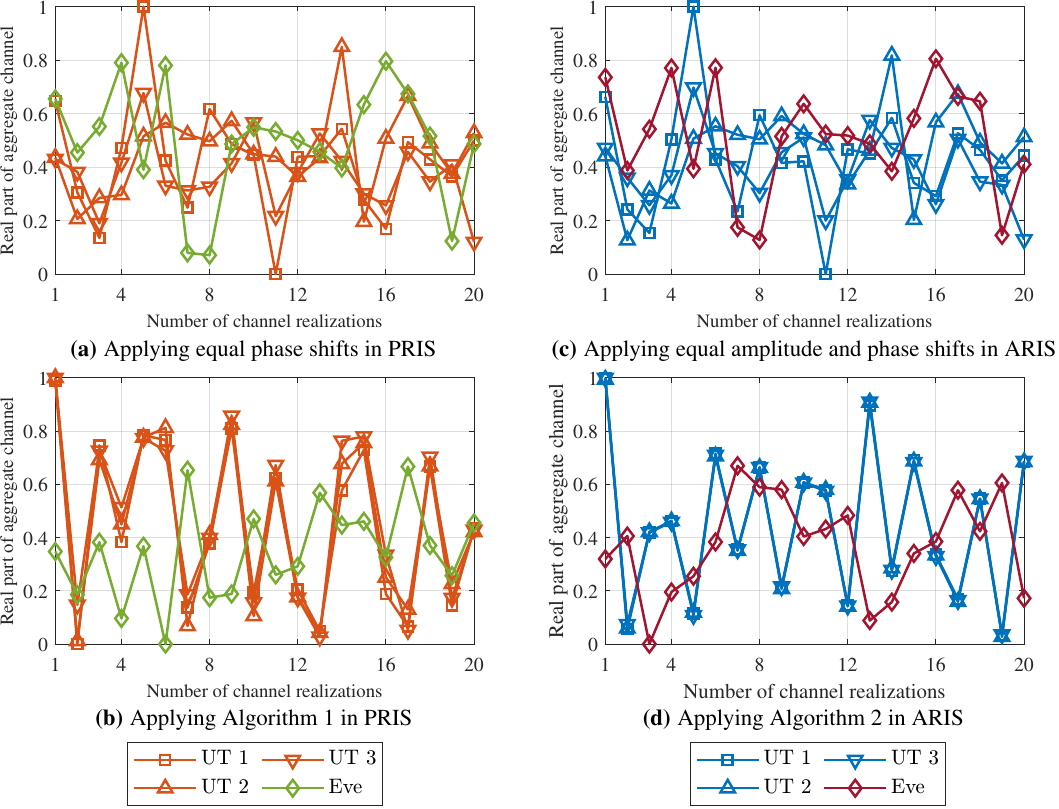}
		\caption{The performance of Eve for PRIS and ARIS.}
		\label{Eve_Performance}
	\end{center}
\end{figure*}
The performance of Eve is closely linked to the correlation between the channel samples of the $K$ legitimate users and those observed by Eve. Here we consider a passive eavesdropping model, which is commonly assumed in key generation schemes. All nodes, including Eve, are positioned at least $\lambda/2$ apart, ensuring that the channels between each pair of nodes are independent \cite{LWC_2023_ARIS}, \cite{STAR-RIS_TVT_2024_1}. Figures \ref{Eve_Performance}(a) and \ref{Eve_Performance}(c) depict the real parts of the aggregate channel samples obtained by the legitimate users and Eve under equal phase shifts in PRIS and equal amplification gains and phase shifts in ARIS, respectively. It is evident that all four nodes observe significantly different channel samples. However, by applying Algorithm \ref{MDM} and Algorithm \ref{ARIS-CDM} in PRIS and ARIS, respectively, the legitimate users are able to obtain identical channel samples. Figures \ref{Eve_Performance}(b) and \ref{Eve_Performance}(d) further illustrate that the channel samples among legitimate users in ARIS are more closely aligned than in PRIS. In both cases, however, the common channel feature among legitimate users remains markedly distinct from that observed by Eve. 
\section{Conclusion}\label{conclusion}
In this contribution, we proposed a novel group physical layer key generation mechanism that leverages optimized RIS parameter design to establish common randomness.  For both PRIS and ARIS scenarios, we formulated corresponding optimization problems and addressed them using SCA and SDR-GR techniques, respectively. Analytical and numerical results demonstrate that the proposed method achieves probing complexity comparable to broadcast-based approaches in the worst case while delivering competitive performance in terms of NMSE, KER, KGR and key randomness. For future research, extending the framework to STAR-RIS and BD-RIS architectures can be appealing. Furthermore, the impact of imperfect CSI and discrete RIS phase/amplitude can be investigated for practical deployment. Integrating MIMO-enabled APs with RIS beamforming design for GKG is an interesting area for further research. Finally, leveraging the AI-driven optimization will further help to reduce the computational burden at the AP.
\bibliographystyle{IEEEtran}
\bibliography{Refrences}
\end{document}